\documentclass[aps,prb,amsmath,boldmath,twocolumn,showpacs]{revtex4-1}
\usepackage{epsfig}
\usepackage[nooneline]{subfigure}
\usepackage{graphicx}
\usepackage{amsmath,amssymb}

\begin{document}
\title{Electronic structure, spin excitations, and orbital ordering in a three-orbital model for iron pnictides.}

\author{Sayandip Ghosh}
\email{sayandip@iitk.ac.in, sayandip.cmt@gmail.com}
\author{Avinash Singh}
\affiliation{Department of Physics, Indian Institute of Technology Kanpur
208016, India} 
\pacs{74.70.Xa, 75.10.Lp, 75.30.Ds, 75.25.Dk}

\begin{abstract}
A three-orbital itinerant-electron model involving $d_{xz}$, $d_{yz}$ and $d_{xy}$ Fe $3d$ orbitals is proposed for iron pnictides towards understanding the ($\pi,0$) ordered magnetism and magnetic excitations in these materials. It is shown that this model at half filling yields a gapped ($\pi,0$) magnetic state with high degree of robustness and stability, and simultaneously reproduces several experimentally observed features such as the electronic structure, spin excitations, as well as the ferro orbital order between the $d_{xz}$ and $d_{yz}$ orbitals.
\end{abstract}
\maketitle

\section{Introduction}
Iron pnictides exhibit a rich temperature-doping phase
diagram \cite{Zhao2008nature,Nandi2010} involving antiferromagnetic (AF),
structural and superconducting phase transitions. Several important microscopic
properties such as electronic Fermi surface structure, orbital order and spin
wave excitations have been extensively investigated experimentally using angle
resolved photoemission spectroscopy (ARPES), x-ray linear dichroism (XLD), and
neutron scattering. From a theoretical point of view, multi-orbital nature of
these materials makes it challenging to understand these properties within a
single theoretical framework.

First principle band-structure calculations \cite{Mazin2008,Haule2008,Singh2008,Nekrasov2008,Zhang2009} have suggested that the density of states (DOS) near Fermi energy is contributed primarily by Fe 3$d$-bands \cite{Singh2008,Nekrasov2008}. These calculations and ARPES experiments \cite{Yi2009,Kondo2010,Yi2011,Brouet2012,Kordyuk2013} have revealed that there are two circular hole pockets around the center ($\Gamma$ point) and two elliptical electron pockets around the corner (M point) of the 2D Brillouin Zone (BZ) in the paramagnetic state. The Fermi surface (FS) goes through a complex multi-orbital reconstruction through the paramagnetic-to-antiferromagnetic transition \cite{Yi2009,Yi2011}. Apart from the FS structure, ARPES \cite{Shimojima2010,Yi2011,Jensen2011} and XLD \cite{Kim2013} experiments have also revealed the existence of ferro orbital order between $d_{xz}$ and $d_{yz}$ Fe orbitals. In the magnetic state, the Fe $d_{yz}$ band is shifted up relative to the $d_{xz}$ band, causing electron density difference between the two orbitals, which may cause structural phase transition \cite{Lv2009}.

In addition to electronic and structural properties, single crystal neutron
scattering experiments have revealed a stripe antiferromagnetic arrangement of
Fe moments in pnictides, corresponding to an in-plane ordering wave vector ${\bf
Q}$=($\pi, 0$). Experimental investigation of spin wave excitations in these
materials has been extensively carried out by inelastic neutron scattering (INS)
measurements \cite{Zhao2008,Ewings2008,Zhao2009,Diallo2009,Ewings2011,Harriger2011}, and the most striking feature is the remarkably high energy scale of magnetic excitations. The spin wave excitations are sharp, highly dispersive,
and extend up to energies of $\sim$ 200 meV with a well-defined maximum at the
ferromagnetic zone boundary (FZB) corresponding to wave vector ${\bf q}$=($\pi,\pi$). The observed spin wave excitations have strong in-plane
anisotropy, with moderate \cite{Zhao2009} to large damping \cite{Harriger2011},
but the excitations do not dissolve into particle hole continuum \cite{Ewings2008}.

Understanding all these magnetic, electronic, and structural properties exhibited by iron pnictides using a single theoretical framework continues to be an outstanding challenge. Although several tight-binding models have been proposed to address these properties, they tend to properly describe either the electronic or the magnetic properties, but not simultaneously both. For example, single band $t-t'$ Hubbard model \cite{Raghuvanshi2010} gives stable ($\pi,0$) state at intermediate hole doping with carrier-induced ferromagnetic (F) spin couplings, correct spin wave dispersion, but does not yield the correct electronic structure. Among the two-orbital models \cite{Raghu2008,Han2008,Daghofer2008}, the minimal two-band model by Raghu \textit{et al.} \cite{Raghu2008} reproduces the FS structure consistent with LDA calculations at half-filling, and nesting between circular hole and electron pockets significantly reduces the critical interaction strength $U_{\rm c}$ for ($\pi,0$) ordering. However, spin wave dispersion in this model does not agree in detail with INS measurements \cite{Raghuvanshi2011,Knolle2011}. Moreover, it does not include $d_{xy}$ Fe orbital which contributes some portions of the electron pockets. Similarly, three-orbital models with one-third filling (i.e. two electrons in three orbitals) \cite{Lee2008}, two-third filling (i.e. four electrons in three orbitals) \cite{Daghofer2010,Zhou2010}, and four-orbital model at half-filling \cite{Yu2009} can reproduce the desired FS structure, but spin wave excitations in these models have not been investigated yet. More realistic five-orbital models \cite{Graser2009,Ikeda2010,Graser2010}, which yield the FS topology similar to experimental findings, were proposed aimed at investigation of pairing instabilities, but spin excitations were not studied. Although anisotropic spin wave excitation was obtained in a recent study \cite{Kaneshita2010} for a five-orbital model \cite{Kuroki2008}, investigation of spin excitations over the entire BZ was not carried out. 

In this context, we present and investigate a three-orbital itinerant-electron model in this paper. We find that this minimal model simultaneously yields the correct FS topology as well as spin wave dispersion consistent with INS experiments. Moreover, ferro orbital order of appropriate sign is also obtained in this model.

The organization of this paper is as follows. The importance of ferromagnetic (F) spin couplings on the experimentally measured spin wave dispersion in the ($\pi,0$) state is briefly discussed in section 2. Then in section 3, it is shown that no F spin coupling is generated in the two-band model with FS nesting \cite{Raghu2008} ,although this model yields correct FS topology. A third $d_{xy}$ Fe orbital is therefore necessary to overcome this shortcoming, and a three-orbital model having $d_{xz}$, $d_{yz}$ and $d_{xy}$ Fe 3$d$ orbitals is presented in section 4 highlighting the Fermi surface and density of states. The investigation of magnetic excitations and orbital ordering in the ($\pi,0$) magnetic state is carried out in section 5. Finally conclusions are discussed in section 6.

\section{Spin wave energy at the ferromagnetic zone boundary and ferromagnetic spin coupling}
\begin{figure}
\hspace{0mm}
\vspace{0mm}
\includegraphics[width=55mm,angle=0]{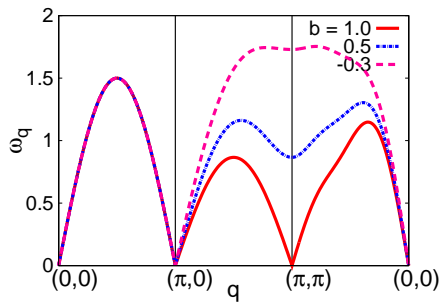}
\caption{\label{wq}
Spin wave dispersion along different BZ directions obtained from Eq. [\ref{magnon dispersion}] for $J'/J = 1$, $J=1/2$, and different values of the coefficient $b$ corresponding to different F spin couplings --- (i) $b=1$ (no F spin coupling), (ii) $b=0.5$ (intermediate case), and (iii) $b<0$ ($J_{\rm F} > J_{\rm AF}$), which yields maximum spin wave energy at the ferromagnetic zone boundary ($\pi,\pi$).
}
\vspace{0mm} 
\end{figure}

In order to highlight the effect of induced F spin couplings on the spin wave dispersion and spin wave energy at the FZB, we consider the case of single band Hubbard model with nearest-neighbor (NN) hopping $t$ and next-nearest-neighbor (NNN) hopping $t'$. In this case, the spin wave dispersion in the strong coupling limit\cite{arxiv2001} is given by,
\begin{eqnarray}
\left ( \frac{\omega_{\bf q}}{J} \right )^2=
\left \{ (1+\frac{2J'}{J})-b(1- \cos q_y) \right \}^2 \nonumber \\ 
- \left \{ (1+\frac{2J'}{J} \cos q_y)\cos q_x \right \}^2. 
\label{magnon dispersion}
\end{eqnarray}
where $J=4t^2/U$ and $J'=4t'^2/U$ represent the NN and NNN AF spin couplings respectively, and the coefficient $b=1$ at half-filling. In presence of doping, carrier-induced F spin coupling are generated,\cite{Raghuvanshi2010} and the spin wave dispersion can be well described by the same expression where, $b \sim (J_{\rm AF} - J_{\rm F})/J_{\rm AF}$ provides a relative measure of the F and AF spin coupling strengths.
The case $b=1$ corresponds to zero F spin couplings, as obtained for the
half-filled $t-t'$ Hubbard model, and yields zero spin wave energy ($\omega_{\rm FZB}$) at the zone boundary in the ferromagnetic direction ($\pi,\pi$). With the onset of F spin couplings ($b < 1$) as in the doped $t-t'$ Hubbard model,\cite{Raghuvanshi2010} the spin wave energy at this wave vector increases from zero  [Fig. \ref{wq}]. When the coefficient $b$ becomes negative at sufficiently large F spin couplings, the spin wave
energy at this wave vector becomes maximum over the entire Brillouin zone,
as indeed is observed in INS measurements on iron pnictides.

\section{Absence of ferromagnetic spin coupling in the two-band nesting model}
We will now investigate spin wave excitations in the minimal two-band model\cite{Raghu2008}  which has been investigated intensively in recent years. The model has two degenerate $d_{xz}$ and $d_{yz}$ orbitals in 2D, and yields nearly circular hole pockets at $(0,0)$, $(\pm \pi,\pm \pi)$ and electron pockets at $(\pm \pi,0)$, $(0,\pm \pi)$. The appreciable FS nesting results in strong instability towards $(\pi,0)$ or $(0,\pi)$ magnetic ordering at significantly low values of the interaction strength ($U_{\rm c} \approx 3$). However, as we will see in this section, the very same critical features for FS nesting (nearly identical circular hole and electron pockets) also necessarily imply vanishing F spin couplings and consequent vanishing of $\omega_{\rm FZB}$, in sharp contrast with INS measurements.
 
The tight-binding part of the Hamiltonian is defined as 
\begin{equation}
 H_0 = -\sum_{{\bf i},{\bf j}}\sum_{\mu,\nu}\sum_{\sigma} t_{{\bf i},{\bf j}}^{\mu,\nu} 
(a_{{\bf i},\mu,\sigma}^\dagger a_{{\bf j},\nu,\sigma} + \text{H.c.}), 
\end{equation} 
where ${\bf i},{\bf j}$ refer to site indices, $\mu,\nu$ are the orbital indices, and  $t_{{\bf i},{\bf j}}^{\mu,\nu}$ are the hopping terms as defined as Ref.  \onlinecite{Raghu2008}. 

\begin{figure}
  \subfigure[]{\includegraphics[height=35mm,width=35mm,angle=-90]{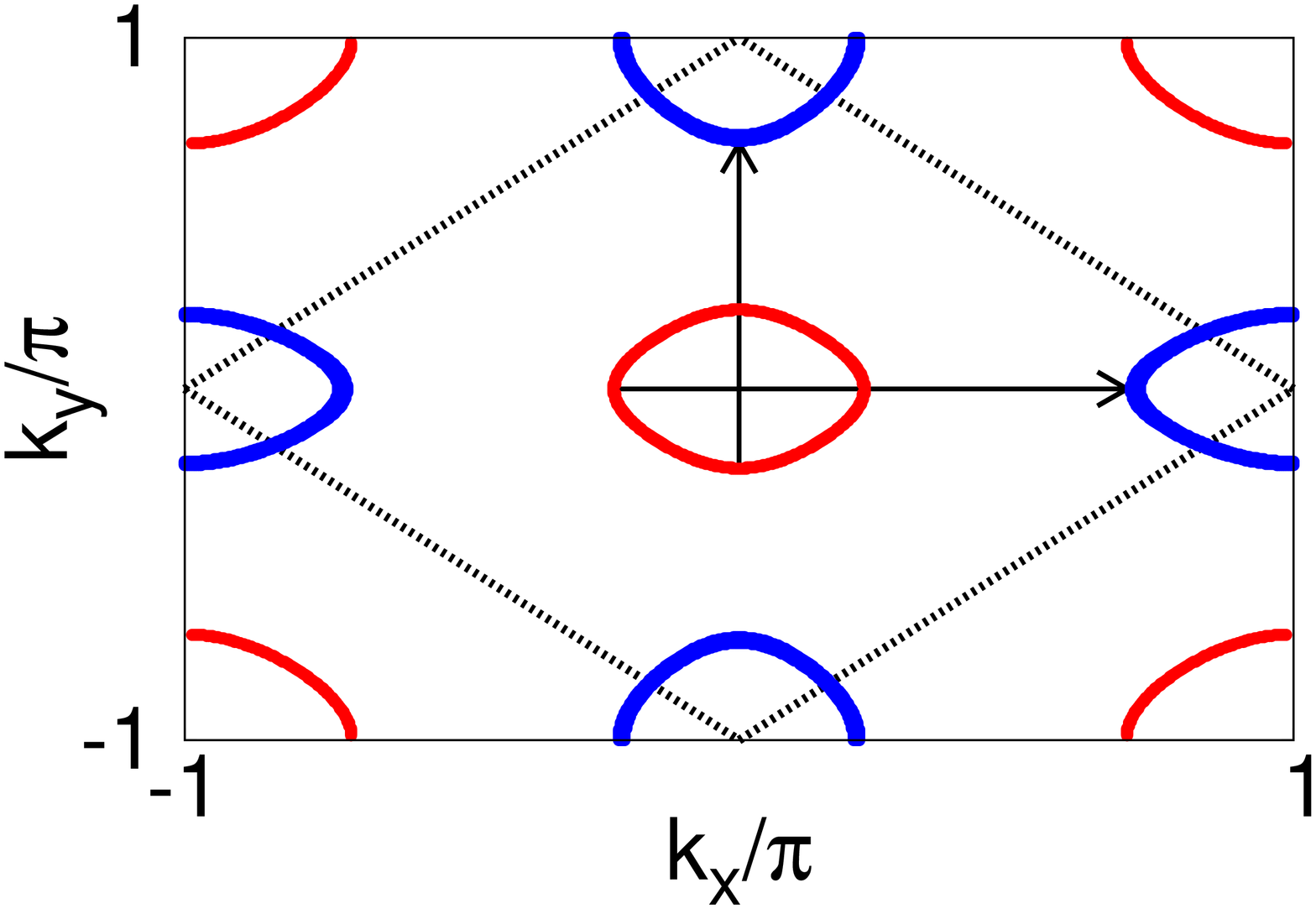}
 \label{fs_twoband1}}
\subfigure[]{\includegraphics[height=40mm,width=40mm,angle=-90]{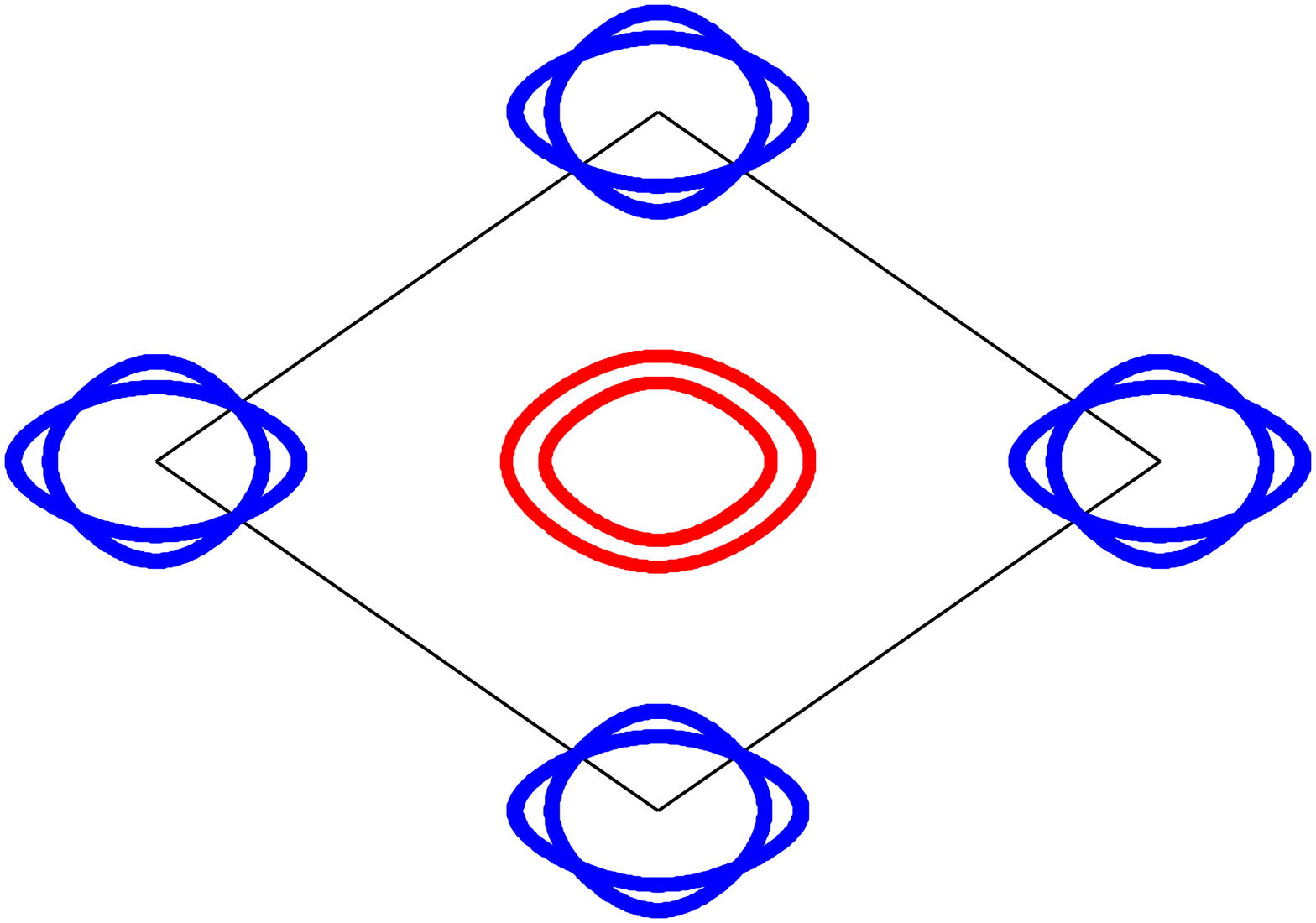}
 \label{fs_twoband2}}
\caption{ \label{fs_twoband} Fermi surface for the half-filled two band model with $t_1=-1.0$, $t_2 = 1.3$, $t_3=t_4=-0.85$ in the \subref{fs_twoband1} unfolded (1 Fe atom/unit cell) BZ and \subref{fs_twoband2} folded BZ. The hole and electron pockets are separated by wave-vector ($\pi,0$) and ($0,\pi$), showing Fermi surface nesting. }
\end{figure}

Figure \ref{fs_twoband} shows the Fermi surface structure for this model. The hopping parameters are $t_1$=$-1.0$, $t_2$=$1.3$, $t_3$=$t_4$=$-0.85$, and Fermi energy $E_F$=$1.45$, in units of $|t_1|$. This Fermi energy corresponds to half-filling (total electron filling $n \sim 2$). In the large BZ (for 1 Fe/cell) [Fig. \ref{fs_twoband1}], there are two hole pockets and two electron pockets. After folding along the faint line in Fig. \ref{fs_twoband1}, FS topology in the actual crystallographic BZ is obtained as in Fig. \ref{fs_twoband2}, which is similar to LDA calculations.

The interaction Hamiltonian consists of different on-site electron-electron interactions:
\begin{eqnarray}
H_I &=& U \sum_{{\bf i},\mu} n_{{\bf i},\mu,\uparrow} n_{{\bf i},\mu,\downarrow} + (U' -
\frac{J}{2}) \sum_{{\bf i},\mu,\nu}^{\mu<\nu} n_{{\bf i},\mu} n_{{\bf i},\nu} \nonumber \\ 
&-& 2 J \sum_{{\bf i},\mu,\nu}^{\mu<\nu} {\bf{S_{{\bf i},\mu}}} \cdot {\bf{S_{{\bf i},\nu}}} \nonumber \\
&+& J' \sum_{{\bf i},\mu,\nu}^{\mu<\nu}
(a_{{\bf i},\mu,\uparrow}^{\dagger}a_{{\bf i},\mu,\downarrow}^{\dagger}a_{{\bf i},\nu,\downarrow}
a_{{\bf i},\nu,\uparrow} + \text{H.c.}),
\label{interaction}
\end{eqnarray}
where ${\bf S}_{i,\mu}$ ($n_{{\bf i},\mu}$) refer to the local spin (charge) density operators. The first and second terms are the intra-orbital and inter-orbital Coulomb interactions respectively, the third term is the Hund's coupling and the fourth term the ``pair-hopping'' term whose coupling strength $J'$ is equal to $J$. From rotational symmetry, $U' = U - 2J$.

The static susceptibility in the paramagnetic state has strong enhancements at ($\pm \pi,0$), ($0,\pm \pi$) due to nesting between hole and electron Fermi pockets,\cite{Raghu2008,Brydon2011} favoring transition to a ($\pi,0$) ordered spin density wave (SDW) state at some low critical interaction strength.

\begin{figure}
\subfigure{\includegraphics[width=55mm,angle=0]{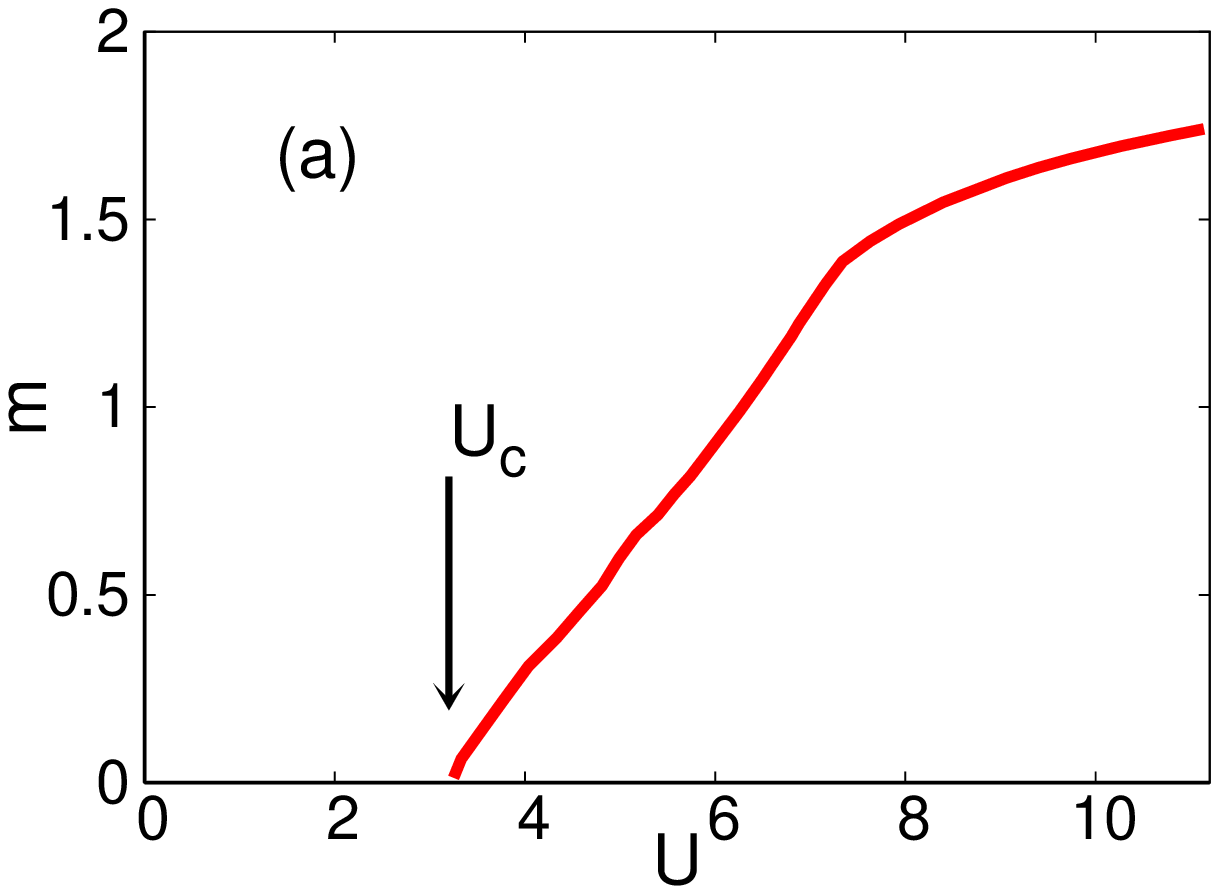}
  \label{nesting a}}
  \subfigure{\includegraphics[width=55mm,angle=0]{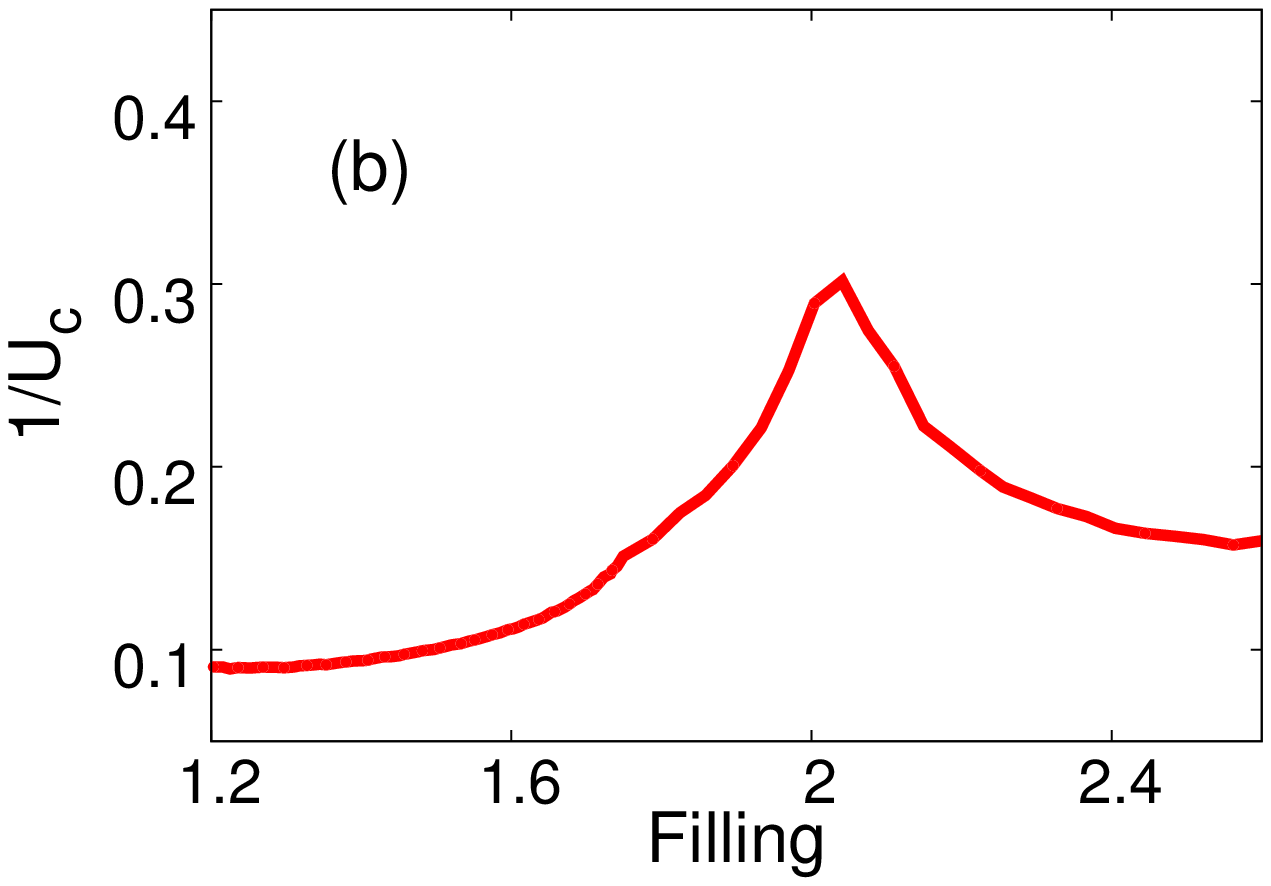}
  \label{nesting b}}
 \caption{\label{Ucr} (a) Variation of sublattice magnetization with $U$ for $n \sim 2$ and $J$=$0$, showing onset of ($\pi,0$) magnetic order at critical interaction strength $U_{c}\approx$ 3.0.  (b) The inverse of critical interaction $U_{c}^{-1}$ shows a maximum around electron filling $n \sim 2$ corresponding to nesting between hole and electron Fermi pockets, showing proclivity of the two-band system towards ($\pi,0$) magnetic ordering. }
\end{figure}

In the ($\pi,0$) state, the Hartree-Fock level (mean-field) Hamiltonian matrix in the composite sublattice-orbital basis (A$xz$ A$yz$ B$xz$ B$yz$) takes the form:
\begin{eqnarray}
&& H_{\rm HF}^{\sigma} ({\bf k}) =  \nonumber \\
&& \left [ \begin{array}{cccc} -\sigma
\Delta_{xz} + \varepsilon_{\bf k}^{2y}   & 0 & \varepsilon_{\bf k}^{1x}
+ \varepsilon_{\bf k}^{3} & \varepsilon_{\bf k}^{4}
\\ 
0 & -\sigma \Delta_{yz} + \varepsilon_{\bf k}^{1y} & \varepsilon_{\bf k}^{4} & \varepsilon_{\bf k}^{2x} + \varepsilon_{\bf k}^{3} \\ 
\varepsilon_{\bf k}^{1x} + \varepsilon_{\bf k}^{3} & \varepsilon_{\bf k}^{4} &
\sigma \Delta_{xz} + \varepsilon_{\bf k}^{2y} & 0 \\
\varepsilon_{\bf k}^{4} & \varepsilon_{\bf k}^{2x} + \varepsilon_{\bf k}^{3} &
0 & \sigma \Delta_{yz} + \varepsilon_{\bf k}^{1y}  \\
\end{array} 
\right ] \nonumber \\
\label{Hamiltonian_twoband}
\end{eqnarray}
for spin $\sigma$ = +(-) for spin $\uparrow$($\downarrow$), where the band energies
\begin{eqnarray}
& & \epsilon_{\bf k}^{1x} = -2 t_1 \cos k_x , \;\;\;\;\;\; \epsilon_{\bf k}^{2x} = -2 t_2 \cos k_x,  \nonumber \\
 & & \epsilon_{\bf k}^{1y} = -2 t_1 \cos k_y,  \;\;\;\;\;\; \epsilon_{\bf k}^{2y} = -2 t_2 \cos k_y,  \\
& & \epsilon_{\bf k}^{3} = -4 t_3 \cos k_x \cos k_y,  \;\;\;\;\;\; \epsilon_{\bf k}^{4} = -4 t_4 \sin k_x \sin k_y, \nonumber
\end{eqnarray}
corresponding to different hopping terms in different directions, and the self-consistently determined exchange fields:
\begin{eqnarray}
2\Delta_{xz} &=& U m_{xz} + Jm_{yz} \nonumber \\
2\Delta_{yz} &=& U m_{yz} + Jm_{xz}
\end{eqnarray}
in terms of the sublattice magnetizations $ m_{xz}$ and $ m_{yz}$ for the two orbitals.  For a given Fermi energy $E_F$,
the sublattice magnetization is obtained from the corresponding electronic
densities as 
\begin{eqnarray}
m_{\mu} &=& n_{\uparrow}^{\mu A} - n_{\downarrow}^{\mu A} = n_{\downarrow}^{\mu
B} - n_{\uparrow}^{\mu B} = n_{\uparrow}^{\mu A} - n_{\uparrow}^{\mu B}
\nonumber \\ 
&=& \sum_{{\bf k},l}[(\phi_{{\bf k},\uparrow,l}^{\mu,A})^2-(\phi_{{\bf
k},\uparrow,l}^{\mu,B})^2] \varTheta(E_F - E_{{\bf k},\uparrow,l}) .
\label{magnetization}
\end{eqnarray}
Here $E_{{\bf k},\sigma,l}$ and $\phi_{{\bf k},\sigma,l}$ are
the eigenvalues and eigenvectors of the Hamiltonian matrix (Eqn.
\ref{Hamiltonian_twoband}), where the index $l$ refers to the four eigenvalue
branches. The pair hopping term does not contribute to the mean field Hamiltonian in
the magnetic state.

Figure \ref{nesting a} shows the evolution of the total sublattice magnetization $m = m_{xz} + m_{yz}$ at half-filling with interaction $U$. Onset of magnetization at $U\approx$ 3.0 indicates the magnetic instability of the system at the critical interaction $U_{\rm c} \approx$ 3.0. Variation of $U_c$ with electron filling [Figure \ref{nesting b}] shows that $U_{c}$ for $(\pi,0)$ ordering has lowest value at half-filling for which the electron and hole pockets are well nested. This highlights the proclivity of the two-band system towards ($\pi,0$) ordering due to FS nesting.

\begin{figure}
\hspace{0mm}
\vspace{0mm}
\includegraphics[width=55mm,angle=0]{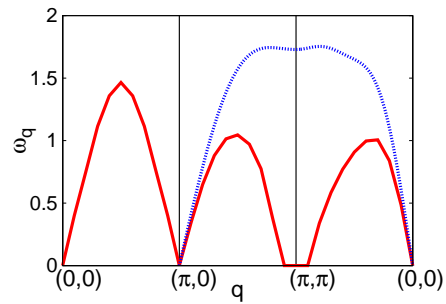}
\caption{\label{wq_twoband}
The spin wave dispersion (red solid line) for the half-filled two-band model with Fermi surface nesting shows identical behavior as for the half-filled $t-t'$ Hubbard model [Fig. \ref{wq}]. The spin wave energy at
the ferromagnetic direction zone boundary ($\pi,\pi$) vanishes, whereas typical INS measurements on iron pnictides show a maximum at this wave vector. The blue dotted line here explicitly shows the importance of F spin coupling for obtaining maximum spin wave energy at FZB as in figure 1 (pink dashed line), and is similar to fitting of experimental data \cite{Zhao2009,Harriger2011}. Here, $\Delta_{xz}$=$\Delta_{yz}$=$2.0 $ for $U \approx 5$ and $J\approx U/4$.
}
\vspace{-0mm} 
\end{figure}

Using eigenvalues and eigenvectors of the Hartree-Fock (HF) level Hamiltonian in the ($\pi,0$) state, we have calculated the spin wave energies, as described in the Appendix A. The calculated spin wave energies along various symmetry directions of the BZ are shown in figure \ref{wq_twoband}. The AF ordering is stable with positive spin
wave energies along AF direction, but spin wave energy at the zone boundary in the ferromagnetic direction vanishes, which is in sharp contrast to the maximum spin wave energy observed at this wave vector in INS experiments \cite{Zhao2009,Harriger2011}. The blue dotted line here explicitly shows the importance of F spin coupling for obtaining maximum spin wave energy at FZB as in figure 1, and thus represents both the theoretical and experimental feature. Thus, while the two band model with circular pockets exhibits high magnetic susceptibility at $n \sim 2$, it does not yield a stable $(\pi,0)$ magnetic state with strong ferromagnetic spin couplings as indicated by INS experiments. The reason for this dichotomy is as below. Maximum spin wave energy at the FZB
requires strong ferromagnetic spin couplings as discussed in Sec. II. According
to band theory of ferro magnetism, the induced F spin coupling is determined by
$\langle \nabla_{\bf k}^2 E_{\bf k} \rangle$ which involves the electron band
curvature. Nearly identical circular hole and electron pockets
yield ferromagnetic spin coupling contributions of similar magnitude but opposite sign, resulting in cancellation and hence no net F spin coupling is induced.

While the strong magnetic response of a system with Fermi surface nesting towards $(\pi,0)$ magnetic ordering may suggest presence of both AF and F spin couplings in respective directions, that is actually not the case. Indeed, the $(\pi,0)$ state can be stabilized by only AF spin couplings, as realized in the half-filled $t-t'$ Hubbard model for $t' > t/2$ in two dimensions \cite{arxiv2001}. While the nearest-neighbor (NN) AF spin couplings get fully frustrated in this magnetic state, the next-nearest-neighbor (NNN) AF spin couplings stabilize the $(\pi,0)$ state which involves AF spin ordering in the diagonal (NNN) directions. That the spin wave dispersion obtained for the two-band model with Fermi-surface
nesting [Figure \ref{wq_twoband}] is similar to that of the half-filled $t-t'$ Hubbard model [Figure \ref{wq}] clearly shows that the $(\pi,0)$ magnetic state in the two-band model is stabilized by only AF spin couplings. Thus, we infer that although nesting condition of the two-band model at half filling strongly enhances the proclivity of the system towards $(\pi,0)$ magnetic ordering, it results in identically vanishing F spin coupling. Hence, it is of interest to include $d_{xy}$ orbital in a minimal model, as indicated by the presence of $d_{xy}$ orbital contribution in the electronic FS structure.

\section{Three-orbital model}
 We therefore consider a half-filled three-orbital model involving $d_{xz}$, $d_{yz}$  and $d_{xy}$ Fe $3d$ orbitals and investigate whether it can simultaneously reproduce essential features of electronic structure as well as the correct spin wave dispersion.

Before we proceed further, it is important to note that the model is defined in the pseudo-crystal momentum ($\bf k$) space, while ARPES experiments probe electronic structure in the physical crystal momentum (${\bf K}$) space. The two momenta are connected by ${\bf K}$=$\bf k$ for $xz$ and $yz$ orbitals, and ${\bf K}$=${\bf k} + {\bf Q}$ for $xy$ orbital (or vice-versa), where ${\bf Q}$=($\pi,\pi$) \cite{Lee2008}. In this ${\bf K}$-space, the discussed two-band model yields one hole pocket at ($0,0$) and one at ($\pi,\pi$), contrary to ARPES measurements.

All iron pnictides have quasi-two dimensional crystal structure with layers of
FeAs stacked along the $c$ axis. The As ions sit alternately above and below the
square plaquettes formed by Fe ions and all interesting phenomena of
superconductivity and magnetism originate from these FeAs layers. We focus on
these FeAs planes and construct a 2D three-orbital model of Fe $d_{xz}$,
$d_{yz}$ and $d_{xy}$ orbitals since these orbitals contribute to the bands near
the Fermi energy.\cite{Mazin2008,Eschring2009,Yi2011}
The orbitals are labeled as $xz$, $yz$, and $xy$ respectively. Hybridization of
Fe 3$d$ orbitals with themselves and through As 3$p$ orbitals contribute to the
effective Fe-Fe hoppings in our model. While $xz$ and $yz$ are degenerate due to
$C_4$ symmetry, $xy$ has a higher onsite energy due to crystal field splitting.

The tight binding Hamiltonian consists of hopping term and onsite energy term
\begin{equation}
 H_0 = -\sum_{{\bf i},{\bf j}}\sum_{\mu,\nu}\sum_{\sigma} t_{{\bf i},{\bf j}}^{\mu,\nu} 
(a_{{\bf i},\mu,\sigma}^\dagger a_{{\bf j},\nu,\sigma} + \text{H.c.}) +    \sum_{\bf i} \sum_{\mu}\varepsilon_{\mu} n_{{\bf i},\mu}, 
\end{equation}
where $a_{{\bf i},\mu,\sigma}^\dagger$ creates an electron at site ${\bf i}$
with spin $\sigma$ in the $\mu$-th orbital, and $t_{{\bf i},{\bf j}}^{\mu,\nu}$
are the hopping amplitudes. We consider NN and NNN hoppings for all the
orbitals. The hopping tensor $t_{{\bf i},{\bf j}}^{\mu,\nu}$ is defined in the
same way as in Ref \onlinecite{Daghofer2010}. $\varepsilon_{\mu}$ is the onsite
energy for the $\mu$-th orbital.

In the plane wave basis defined as $a_{{\bf i},\mu,\sigma} = \frac{1}{\surd
N}\sum_{\bf k}e^{i{\bf k}\cdot{\bf r}_i}a_{{\bf k},\mu,\sigma}$, the
tight-binding Hamiltonian is given by
\begin{equation}
 H_0 = \sum_{\bf k} \sum_{\sigma} \sum_{\mu,\nu} T^{\mu,\nu}({\bf k}) a_{{\bf
k},\mu,\sigma}^{\dagger} a_{{\bf k},\nu,\sigma},
\label{tight}
\end{equation}
where
\begin{eqnarray}
T^{11} &=& - 2t_1\cos  k_x - 2t_2\cos  k_y - 4t_3 \cos  k_x 
\cos  k_y \label{eq:t11}\nonumber \\
T^{22} &=& - 2t_2\cos  k_x -2t_1\cos  k_y - 4t_3 \cos  k_x 
\cos  k_y \label{eq:t22}\nonumber \\
T^{33} &=& - 2t_5(\cos  k_x+\cos  k_y)  - 4t_6\cos  k_x\cos  k_y \nonumber \\ 
&+& \varepsilon_{\rm diff}
\label{eq:t33} \nonumber \\
T^{12} &=& T^{21} =- 4t_4\sin  k_x \sin  k_y \label{eq:t12}\nonumber \\
T^{13} &=& \bar{T}^{31} = - 2it_7\sin  k_x - 4it_8\sin  k_x \cos  k_y
\label{eq:t13} \nonumber \\
T^{23} &=& \bar{T}^{32}= - 2it_7\sin  k_y - 4it_8\sin  k_y \cos  k_x\;
\label{eq:t23}
\end{eqnarray}
are the tight-binding matrix elements in the unfolded BZ ($-\pi \leq k_x,k_y
\leq \pi$). The crystallographic BZ is folded ($|k_x| + |k_y| \leq \pi$) and
folding doubles the number of bands to six. Here, $t_1$ and $t_2$ are the
intra-orbital hopping for $xz$ ($yz$) along $x$($y$) and $y$($x$) directions
respectively; $t_3$ is the intra-orbital hopping along diagonal direction for
$xz$ and $yz$; $t_4$ inter-orbital hopping between $xz$ and $yz$; $t_5$ and
$t_6$ are intra-orbital NN and NNN hoppings for $xy$; $t_7$ and $t_8$ the NN and
NNN hybridization between $xy$ and $xz/yz$. Finally, $\varepsilon_{\rm diff}$ is
the energy difference between the $xy$ and degenerate $xz/yz$ orbitals.

\begin{figure}
 \subfigure[]{\includegraphics[height=50mm,angle=-90]{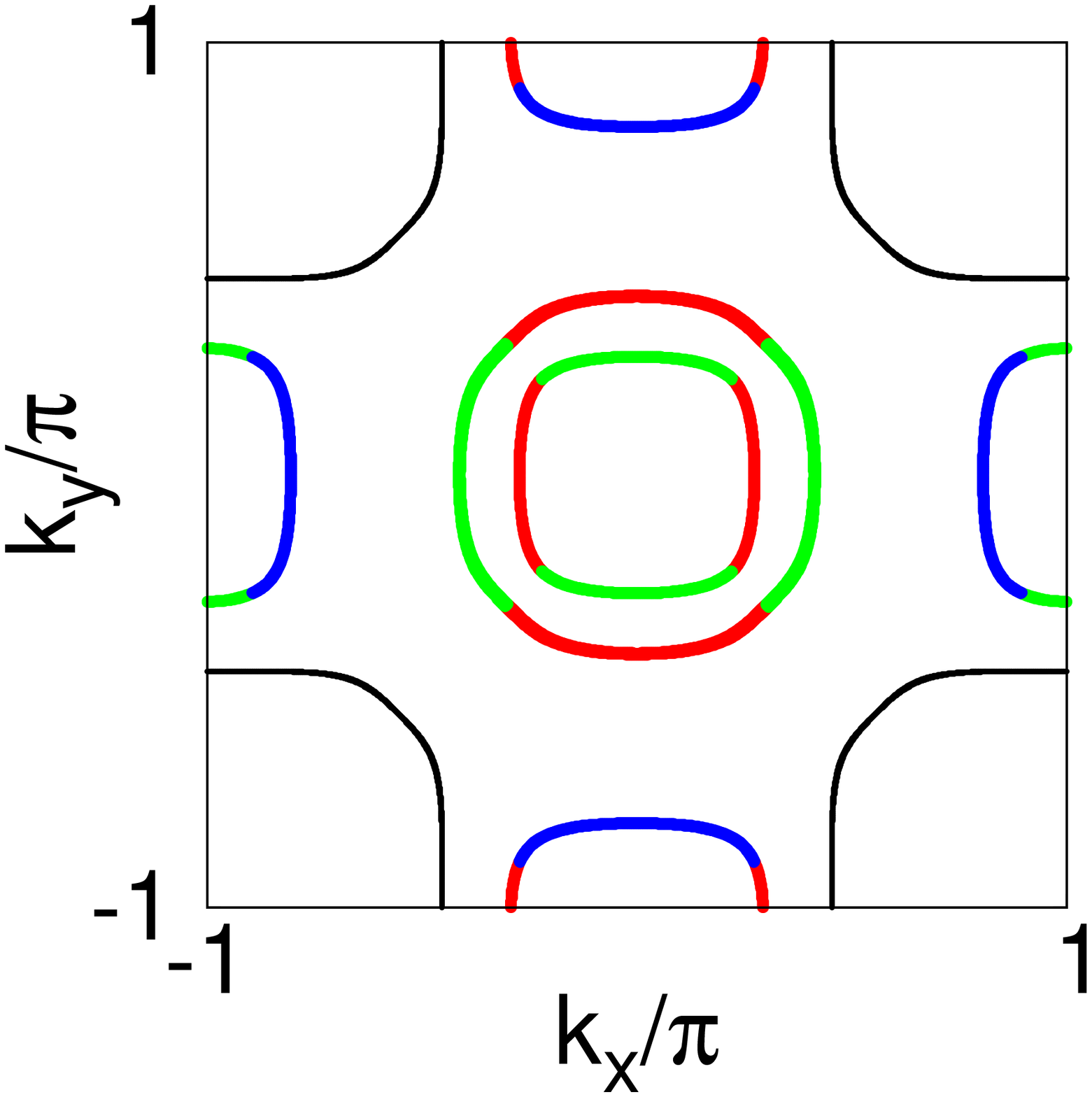}
 \label{fs1}}\\
\subfigure[]{\includegraphics[height=25mm,angle=0]{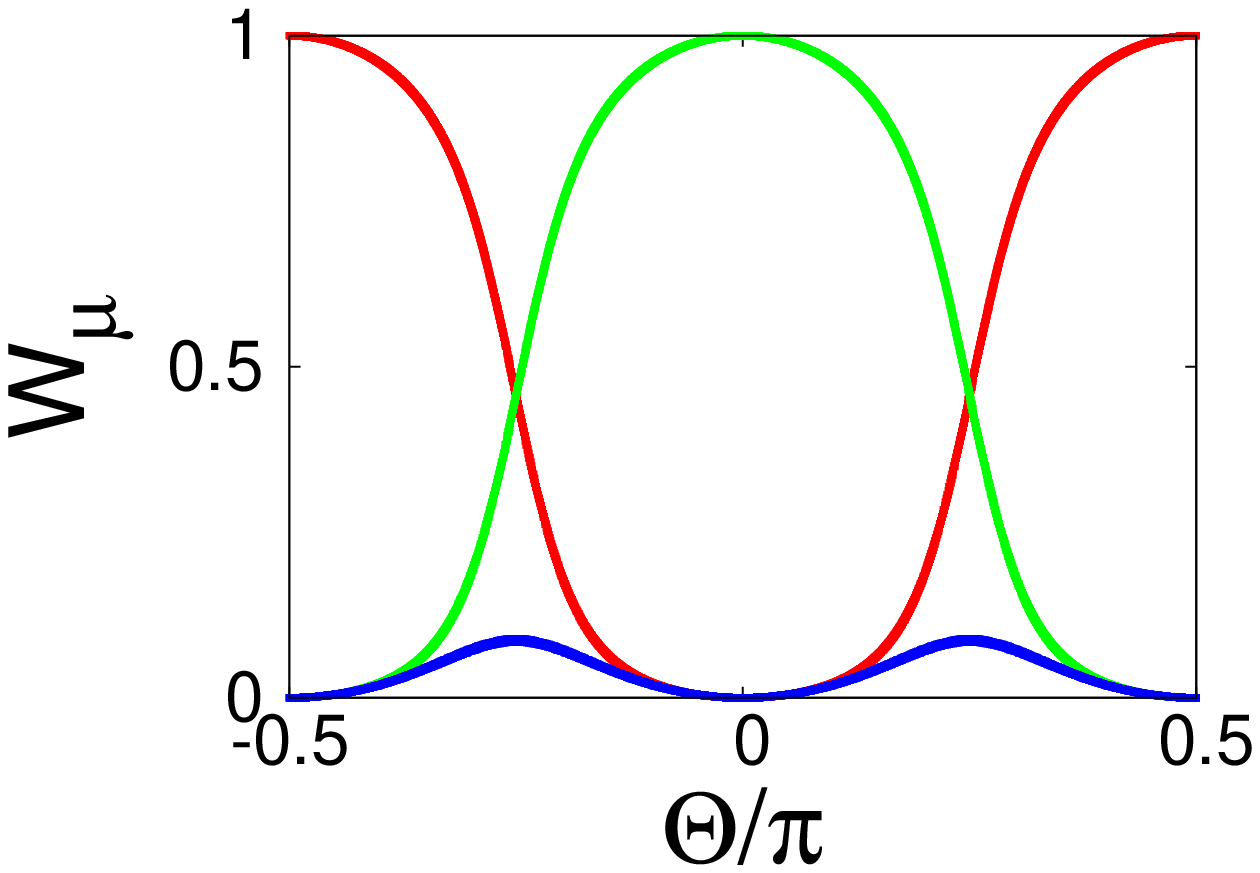}
 \label{fs2}}
  \subfigure[]{\includegraphics[height=25mm,angle=0]{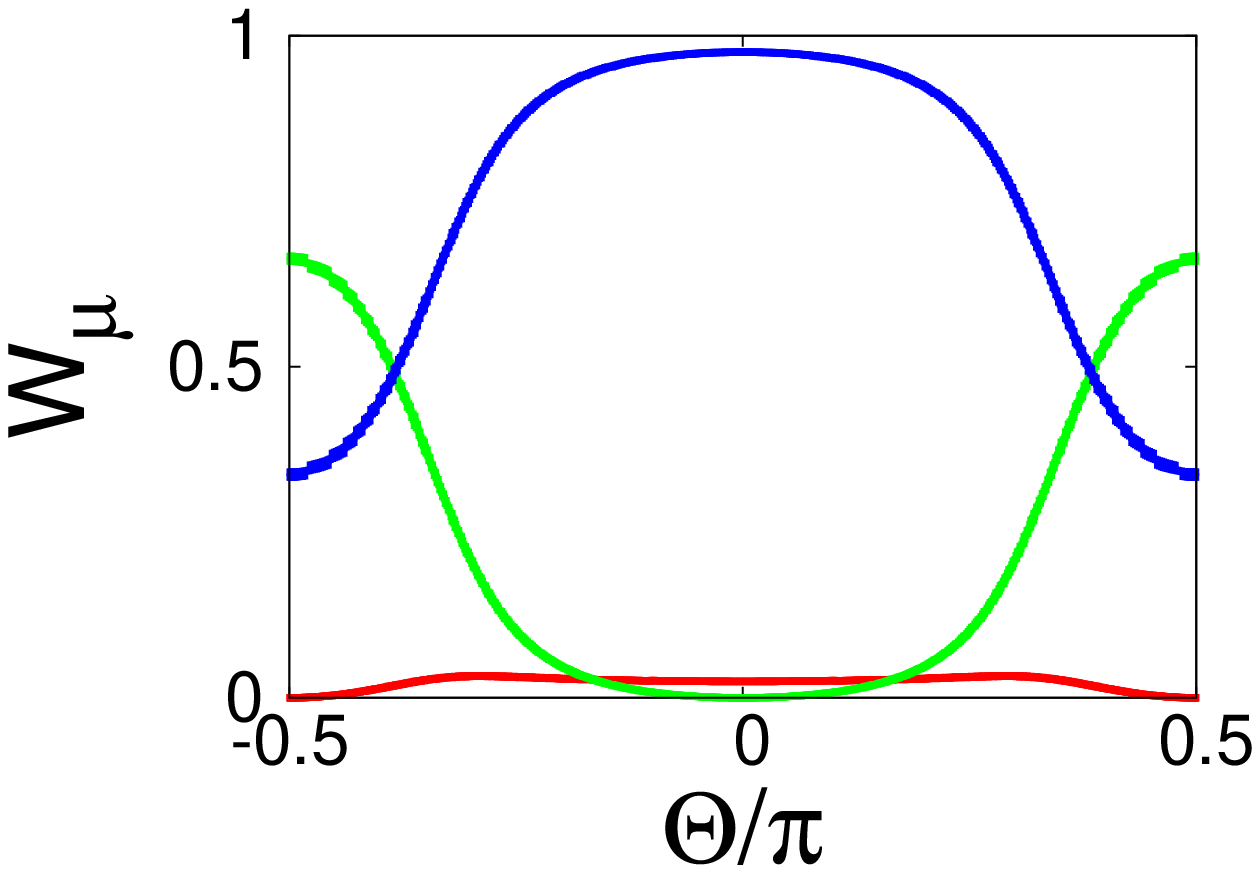}
 \label{fs3}}
 \caption{ \label{fs}\subref{fs1} Fermi surface in the three-orbital model with hopping parameters as given in Table \ref{hoppings} in the unfolded BZ . The main orbital contributions in (a) are shown by following colors: $d_{xz}$ (red), $d_{yz}$ (green), and $d_{xy}$ (blue). Orbital-weight $W_{\mu}$ for the outer hole pocket and electron pocket at ($\pi,0$) in (a) are shown in (b) and (c) respectively. The angle $\Theta$ is measured from $k_x$-axis.
 }
\end{figure}

 \begin{table}
\caption{Value of hopping parameters in the three-orbital model in eV. }
\addtolength{\tabcolsep}{5pt}
 \begin{tabular}{c c c c c c c c c}\hline 
$t_1$ & $t_2$ & $t_3$ & $t_4$ & $t_5$ & $t_6$ & $t_7$ & $t_8$ \\ \hline
 $0.1$ & $0.32$ & $-0.29$ & $-0.06$ & $-0.3$ & $-0.16$ & $-0.15$ & $-0.02$ \\ \hline 
 \end{tabular}
 \label{hoppings}
\end{table}

\subsection{Fermi Surface}
The Fermi surface for the three-orbital tight-binding Hamiltonian in Eq. \ref{tight} is shown in Fig \ref{fs} with the hopping parameter values as listed in Table \ref{hoppings}. This set of hopping parameters will be used throughout this work. Here, Fermi energy and energy difference for $d_{xy}$ are chosen to be $E_F$=$-0.06$ eV and $\varepsilon_{\rm diff}$=$0.32$ eV respectively for half-filling ($n \sim 3$). As shown in Fig. \ref{fs1}, there are two nearly circular hole pockets around the center ($0,0$), and elliptical electron pockets each around ($\pm \pi,0$) and ($0,\pm \pi$) in the unfolded BZ. The two hole pockets are of ${xz}-{yz}$ character, while the electron pocket at ($\pi,0$) [($0,\pi$)] involves hybridization of $xy$ with $yz$ [$xz$] with mainly $xy$ character along $k_x$ [$k_y$] direction. The angular dependence of orbital-weights for the outer hole pocket and the ($\pi,0$) electron pocket are shown in figure \ref{fs2} and \ref{fs3} respectively. The pockets and their orbital content are in good agreement with more realistic five-orbital models \cite{Graser2009,Graser2010,Kuroki2008} and ARPES studies. There is an additional structure around ($\pi,\pi$) originating from the strongly dispersive part of the band falling from the peak at ($\pi,\pi$), and can be easily removed by including a weak third-neighbor hopping terms as included in \cite{Graser2010}.

Thus we see that our three-orbital model can reproduce the essential features of FS structure found out in LDA 
calculations and observed in ARPES experiments. Similar FS topology was also obtained in the two-band model \cite{Raghu2008}. However, the present work provides substantial improvement over this FS structure with regard to nature of pockets and their orbital content. While two-band model yields two nearly similar hole pockets, one each around the ($0,0$) point and ($\pi,\pi$) point in the unfolded BZ, not in agreement with ARPES results, our model yields two hole pockets around the ($0,0$) point. Also, contrary to two-band model which gives circular electron pockets of $xz-yz$ character, our model gives elliptical electron pockets with contributions from $xy$ orbital. This is important for the model to yield two electron pockets as seen in ARPES. Moreover, ellipticity of the electron pockets plays a key role in the spin wave dispersion, as discussed later.

\begin{figure}
\hspace{0mm}
\vspace{0mm}
\includegraphics[width=55mm,angle=0]{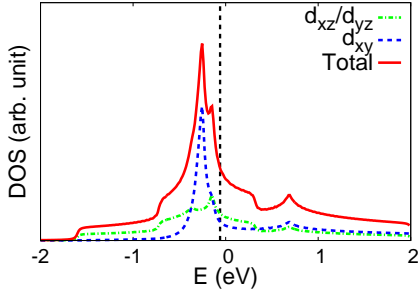}
\caption{\label{dos}
 The partial and total density of states (DOS) in the three-orbital model showing the Fermi energy (dashed line) near the Van Hove singularity which comes mainly from the quasi one-dimensional character of the $d_{xy}$ band.
}
\vspace{-0mm} 
\end{figure}

\subsection{Density of States}
The partial and total density of states (DOS) for the three-orbital model is shown in Fig \ref{dos}. It has a Van Hove singularity just below the Fermi energy which agrees well with LDA results \cite{Haule2008,Singh2008}. In contrast to the two-band model where the Fermi energy lies in the flat region of the band, in our model it is in the vicinity of the Van Hove singularity. This singularity comes primarily from the $d_{xy}$ orbital. Elliptical electron pocket corresponds to quasi-one dimensional character, which has a DOS peak near the lower band edge. Consequently, $\langle \nabla_{\bf k}^2 E_{\bf k} \rangle$ is large which favors strong F spin coupling according to band theory of ferromagnetism. Thus, the ellipticity of electron Fermi pockets in our model plays a pivotal role in generating strong F spin coupling and therefore maximum spin wave energy at ($\pi,\pi$), as investigated in the following section.

\section{($\pi,0$) SDW state in three-orbital model}
\subsection{Magnetic excitations}
In this subsection we will study magnetic excitations in the ($\pi,0$) magnetic state in the three-orbital model introduced in the last section. We will include various electron-electron interaction terms as described in Eq. \ref{interaction}. As the intermediate-coupling regime will be considered throughout, the term “SDW state” is used here without any implicit weak-coupling connotation.

Performing Hartree-Fock approximation as described in Appendix B, HF level (mean-field) Hamiltonian matrix in this composite sublattice-orbital basis (A$xz$ A$yz$ A$xy$ B$xz$ B$yz$ B$xy$) is obtained as:
\begin{widetext}
\begin{equation}
H_{\rm HF}^{\sigma} ({\bf k}) = \left [ \begin{array}{cccccc} -\sigma
\Delta_{xz} + \varepsilon_{\bf k}^{2y}   & 0 & 0 & \varepsilon_{\bf k}^{1x}
+ \varepsilon_{\bf k}^{3} & \varepsilon_{\bf k}^{4} & \varepsilon_{\bf k}^{7x}
+ \varepsilon_{\bf k}^{8,1}
\\ 
0 & -\sigma \Delta_{yz} + \varepsilon_{\bf k}^{1y} & \varepsilon_{\bf k}^{7y}  &
\varepsilon_{\bf k}^{4} & \varepsilon_{\bf k}^{2x} + \varepsilon_{\bf k}^{3}  &
\varepsilon_{\bf k}^{8,2} \\ 
0 & -\varepsilon_{\bf k}^{7y} & - \sigma \Delta_{xy} + \varepsilon_{\bf
k}^{5y} + \tilde{\varepsilon}_{\rm diff} & - \varepsilon_{\bf k}^{7x}- \varepsilon_{\bf
k}^{8,1} & -\varepsilon_{\bf k}^{8,2} & \varepsilon_{\bf k}^{5x} +
\varepsilon_{\bf k}^{6} \\ 
\varepsilon_{\bf k}^{1x} + \varepsilon_{\bf k}^{3} & \varepsilon_{\bf k}^{4} &
\varepsilon_{\bf k}^{7x} + \varepsilon_{\bf k}^{8,1} & \sigma \Delta_{xz} +
\varepsilon_{\bf k}^{2y} & 0 & 0 \\
\varepsilon_{\bf k}^{4} & \varepsilon_{\bf k}^{2x} + \varepsilon_{\bf k}^{3} &
\varepsilon_{\bf k}^{8,2} & 0 & \sigma \Delta_{yz} + \varepsilon_{\bf k}^{1y} &
\varepsilon_{\bf k}^{7y} \\
- \varepsilon_{\bf k}^{7x} - \varepsilon_{\bf k}^{8,1} & -\varepsilon_{\bf
k}^{8,2} & \varepsilon_{\bf k}^{5x} + \varepsilon_{\bf k}^{6} & 0 &
- \varepsilon_{\bf k}^{7y} & \sigma
\Delta_{xy} + \varepsilon_{\bf k}^{5y} + \tilde{\varepsilon}_{\rm diff}  \\
\end{array}
\right ]
\label{Hamiltonian}
\end{equation}
\end{widetext}
for spin $\sigma$, where
\begin{eqnarray}
\varepsilon_{\bf k}^{1x} &=& -2 t_1 \cos k_x  \;\;\;\;\;\;
\varepsilon_{\bf k}^{1y} = -2 t_1 \cos k_y  \nonumber \\
\varepsilon_{\bf k}^{2x} &=& -2 t_2 \cos k_x  \;\;\;\;\;\; 
\varepsilon_{\bf k}^{2y} = -2 t_2 \cos k_y  \nonumber \\
\varepsilon_{\bf k}^{5x} &=& -2 t_5 \cos k_y  \;\;\;\;\;\; 
\varepsilon_{\bf k}^{5y} = -2 t_5 \cos k_y  \nonumber \\
\varepsilon_{\bf k}^{3} &=& -4 t_3 \cos k_x \cos k_y  \;\;\;\;\;\;
\varepsilon_{\bf k}^{4} = -4 t_4 \sin k_x \sin k_y \nonumber \\ 
\varepsilon_{\bf k}^{6} &=& -4 t_6 \cos k_x \cos k_y \nonumber \\
\varepsilon_{\bf k}^{7x} &=& -2i t_7 \sin k_x  \;\;\;\;\;\;
\varepsilon_{\bf k}^{7y} = -2i t_7 \sin k_y  \nonumber \\
\varepsilon_{\bf k}^{8,1} &=& -4i t_8 \sin k_x \cos k_y  \;\;\;\;\;\;
\varepsilon_{\bf k}^{8,2} = -4i t_8 \cos k_x \sin k_y \nonumber \\ 
\end{eqnarray}
are the band energies corresponding to different hopping terms along different directions, and the self-consistent exchange fields are defined as
$2\Delta_\mu = U m_{\mu} + J\sum_{\nu \neq \mu}m_{\nu}$ in terms of sublattice magnetizations $m_{\mu}$ obtained from the corresponding electronic densities as given in Eq. \ref{magnetization}.

It should be noted that, apart from a constant term, the density term $(5J-U) n_\mu /2$ (for orbital $\mu$) arising in the HF approximation has not been shown explicitly in Eqn. \ref{Hamiltonian}. Although the magnitude of this term can be appreciable, only the relative energy shifts $(5J-U)(n_\mu - n_\nu)/2$ are important at fixed filling. For our choice of parameters, the shift between $d_{xz}$ and $d_{yz}$ orbital is very small ($\sim$ 0.03 $U$), and is neglected in our calculations. Although the shift between $d_{xy}$ and $d_{xz/yz}$ orbitals is appreciable, it has been absorbed in the renormalized energy difference $\tilde{\varepsilon}_{\rm diff}$ which has two components $-$ (i) the orbital energy difference $\varepsilon_{\rm diff}$, and (ii) the relative energy shift due to density term in the HF approximation.

Spin-wave energies in this spontaneously broken-symmetry SDW state are calculated as described in the Appendix A. Negative spin wave energies are taken as signature of instabilities of the SDW state. INS experiments measure the dynamical spin structure factor which is proportional to the imaginary part of the transverse spin susceptibility. The transverse susceptibility is calculated by taking trace of the susceptibility matrix [Eq. \ref{chi}]. All our calculations are done at zero temperature.

\begin{figure}
  \subfigure[]{\includegraphics[width=55mm,angle=0]{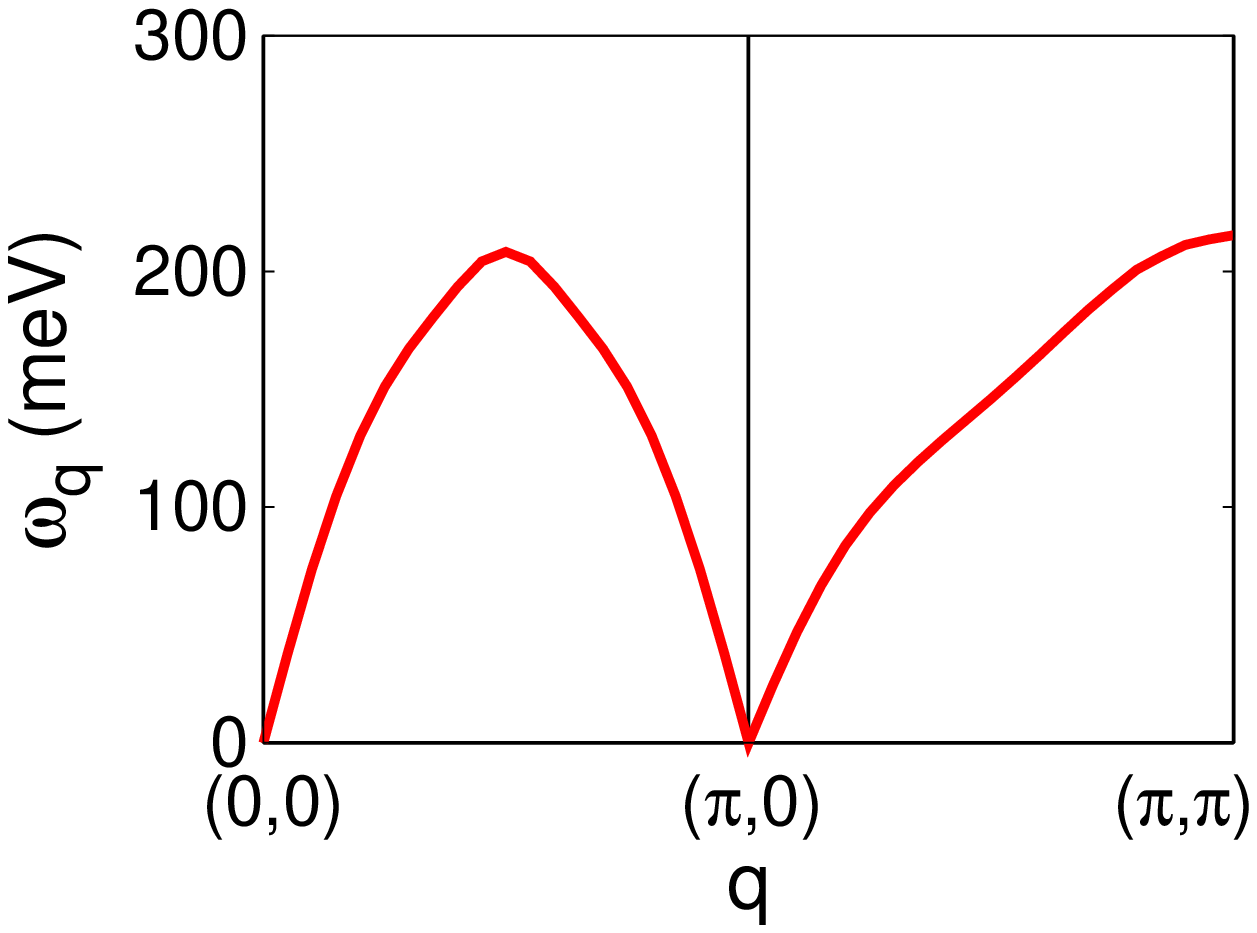}
  \label{spin wave}}
  \hspace{2mm}
  \subfigure[]{\includegraphics[width=65mm,angle=0]{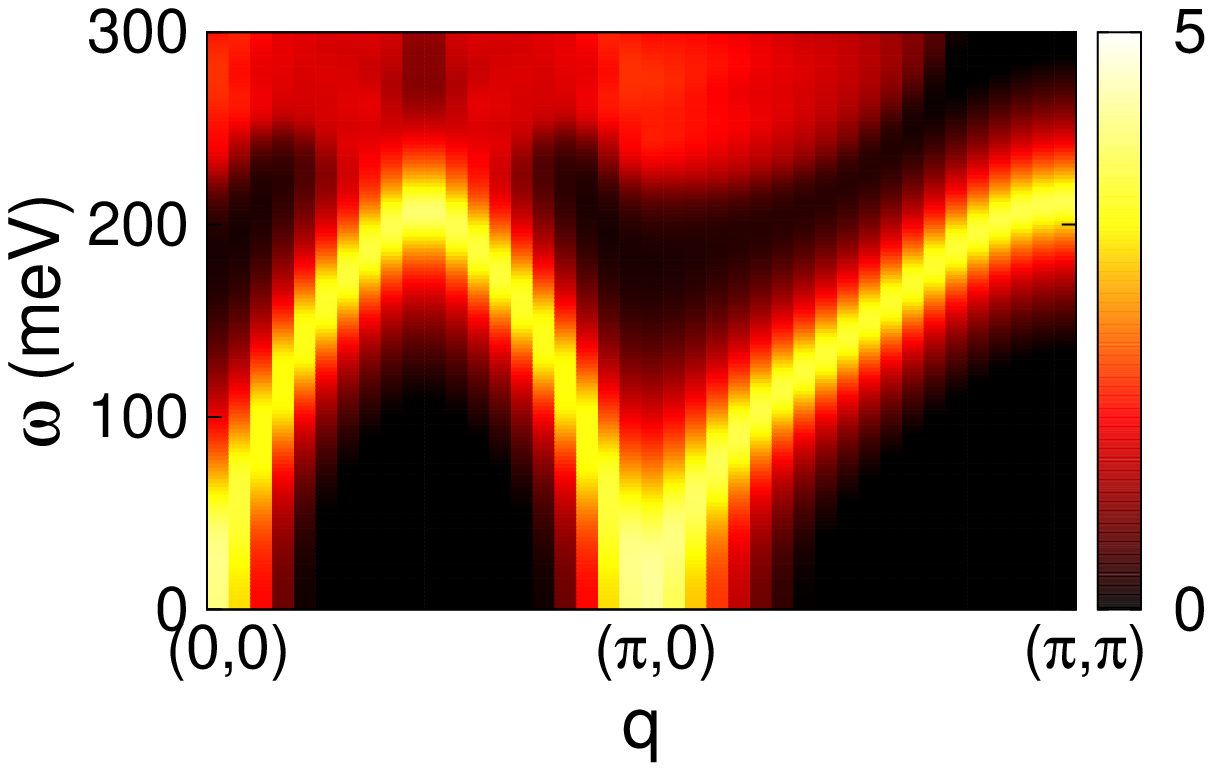}
  \label{sus}}
 \caption{(a) Spin wave dispersion in the ($\pi,0$) SDW state of the three-orbital model with hopping parameters as given in Table \ref{hoppings} . The maximum energy around ($\pi,\pi$) is consistent with INS experiments. (b) Imaginary part of the transverse susceptibility for same parameters.}
\end{figure}

Figure \ref{spin wave} shows spin wave dispersion for ($\pi,0$) SDW state in our three-orbital model. For $U\approx 1.2$ eV, $J\approx U/4$ and total filling $n\approx 3.0$, the self-consistent exchange fields are $\Delta_{xz}\approx 0.53$ eV, $\Delta_{yz}\approx 0.61$ eV, and $\Delta_{xy}\approx 0.57$ eV. The Fermi energy ($E_F$) and renormalized energy difference ($\tilde{\varepsilon}_{\rm diff}$) values are $0.25$ and $0.8$ eV respectively. We see that that not only is the ($\pi,0$) SDW state stable throughout the BZ in our model, as inferred from the positive spin wave energy throughout BZ, the calculated spin wave energy scale is $\sim$ 200 meV which matches well with INS experiments. Most importantly, strong ferromagnetic spin coupling is also generated depicted by the maximum spin wave energy at FZB, consistent with the spin wave dispersion in INS experiments. The values of the interaction strength $U$ and $J$ are realistic and obey the constraints imposed by ARPES and INS experiments on coupling strength for iron pnictides.\cite{Luo2010}

Figure \ref{sus} shows the imaginary part of the transverse spin susceptibility, Im $[\chi^{-+} _{\rm RPA} ({\bf q},\omega)]$. The intensity is shown on a log scale. Evidently, the spin wave excitations are highly dispersive with finite damping. The closed structure of spin excitations over the entire BZ implies that, in contrast to other itinerant models,\cite{Knolle2010,Knolle2011} the magnetic excitations in our model do not rapidly dissolve into particle-hole continuum, as indeed not observed in experiments up to energies $\sim$ 200 meV.

The maximum spin wave energy at the FZB implies strong F spin coupling and ($\pi,0$) SDW state is stabilized without any frustrating NN AF coupling. The ($\pi,0$) state has been shown to be stable also within a spin-only Heisenberg model with comparable NN ($J_1$) and NNN ($J_2$) antiferromagnetic superexchanges.\cite{Yao2008} But such a model is strongly frustrated and can not explain the observed maximum of spin wave dispersion at ($\pi,\pi$). In fact, it was shown that spin wave dispersion throughout the BZ and the maximum at ($\pi,\pi$) can be explained by a suitably parameterized Heisenberg Hamiltonian with an effective ferromagnetic exchange interaction ($J_{1b}<0$) along $b$ direction.\cite{Zhao2009} The present work provides the microscopic origin of this ferromagnetic interaction as due to usual particle-hole exchange mediated spin interaction in an itinerant-electron model, with emergence of strong F spin coupling originating from elliptical nature of electron pockets reflecting quasi-one dimensional electron band.

The magnetic excitation spectrum in our model satisfies Goldstone mode condition due to spin rotational invariance of the model Hamiltonian and yields zero spin wave energy at $\bf q$=($0,0$) and ($\pi,0$). In the low temperature phase for iron pnictides, Fe moments allign along $a$ direction suggesting the need to include single-ion magnetic anisotropy in our model. In fact, orthorhombic distortion can introduce such anisotropy through magnetoelastic coupling and break the rotational invariance. The resulting excitation spectrum will have a spin wave gap as measured in experiments.\cite{Ewings2008,Zhao2008} However, the magnitude of the measured gap was found to be much smaller ($\leq$ 10 meV) than typical spin wave energy scale and can be neglected.

\subsection{Orbital ordering}

\begin{figure}
\hspace{0mm}
\vspace{0mm}
\includegraphics[width=65mm,angle=0]{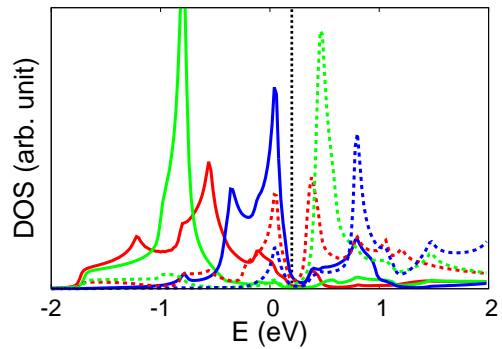}
\caption{\label{dos2}
 The partial density of states (DOS) in the SDW state in the three-orbital model. The color notation is same as in fig. 5. The solid and dashed lines are for majority and minority spins respectively.
}
\vspace{-0mm} 
\end{figure}

As mentioned earlier, ARPES and XLD experiments have found the existence of ferro orbital order between the $d_{xz}$ and $d_{xz}$ orbitals in pnictides. This type of orbital ordering was previously proposed \cite{Lv2009,Lee2009,Chen2010,Lv2011} to explain experimentally observed in-plane anisotropic behavior like anisotropy in magnetic exchange coupling \cite{Zhao2009}, transport properties \cite{Tanatar2010,Chu2010}, FS structure \cite{Yi2011}, and electronic structure \cite{Chuang2010}. 

Figure \ref{dos2} shows the partial density of states in the ($\pi,0$) SDW state with parameters same as in figure 7. While the $d_{xz}$ and $d_{yz}$ orbitals are degenerate in the non-magnetic state [Fig. 6] due to $C_4$ symmetry, the degeneracy is lifted in the ($\pi,0$) state, causing electron density difference in the two orbitals. In the ($\pi,0$) state with exchange field parameters same as used in figure \ref{spin wave}, electron fillings in different orbitals are obtained as $n_{xz}\approx$ 1.2, $n_{yz}\approx$ 1.0 and $n_{xy} \approx$ 0.8. This sign of ferro-orbital order ($n_{xz} > n_{yz}$) is in agreement with experiments \cite{Shimojima2010,Yi2011,Jensen2011,Kim2013}. As highlighted earlier \cite{Ghosh2014}, the presence of hopping anisotropy ($|t_1|<|t_2|$) along with anisotropic ($\pi,0$) magnetic order breaks the equivalence between $a$ and $b$ directions, and  naturally leads to orbital ordered state. A recent  first-principle study \cite{Sen2014} of orbital-dependent electronic structure using experimental inputs have also confirmed this kind of degeneracy-lifting between $d_{xz}$ and $d_{yz}$ orbitals in the magnetic state. 

Spin wave excitations in similar ferro orbital ordered state was previously carried out within a degenerate double-exchange model including antiferromagnetic superexchange interactions \cite{Lv2010}. However, the sign of the reported ($n_{yz} > n_{xz}$) ferro orbital order does not agree with experiments. Furthermore, for a realistic NN hopping value of 200 meV, their calculated spin wave energy scale of around 30 meV is well below the nearly 200 meV energy scale measured in INS experiments.

\subsection{Gapped ($\pi,0$) SDW state}

\begin{figure}
\hspace{0mm}
\vspace{0mm}
\includegraphics[width=65mm,angle=0]{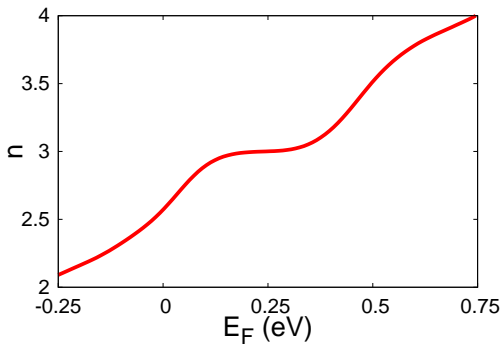}
\caption{\label{fill}
 The total electron filling in the SDW state as a function of the Fermi energy showing a band gap at half filling.
}
\vspace{-0mm} 
\end{figure}

Figure \ref{fill} shows the variation in total electron filling with Fermi energy in the SDW state for same parameters as in figure \ref{spin wave}, clearly indicating presence of an energy gap at half filling, as also seen in figure \ref{dos2} for the SDW state DOS. Moreover, the gapped SDW state shows strong robustness with respect to small variations in hopping parameters, energy difference for $xy$ ($\tilde{\varepsilon}_{\rm diff}$), and interaction strength. 

The hybridization of $xy$ orbital with $xz/yz$ orbitals, which is reflected in the elliptical electron pocket in the FS structure, is also responsible for the energy gap formation in the SDW state, as seen by the absence of energy gap for zero hybridization. Thus, the orbital mixing plays a key role in the formation of the gapped ($\pi,0$) SDW state in the half-filled three-orbital model.

\section{Conclusions}
The minimal two-band model for iron pnictides with nearly circular
electron and hole pockets shows proclivity towards $(\pi,0)$ SDW
ordering due to FS nesting. However, due to cancellation of induced
ferromagnetic spin couplings from electron and hole pocket contributions,
the very same feature of nearly identical electron and hole pockets was shown to yield vanishing ferromagnetic spin coupling
and consequently vanishing spin-wave energy at the ferromagnetic zone
boundary. This being in sharp contrast to the observed maximum spin-wave
energy at this wave vector in inelastic neutron scattering studies necessitated the inclusion of a third $d_{xy}$ orbital to properly account for the experimentally observed spin dynamics in iron pnictides.

A minimal three-orbital itinerant-electron model for iron pnictides involving the  $d_{xz}$, $d_{yz}$, and $d_{xy}$ Fe orbitals was therefore proposed in this paper. Our investigations show the existence of a stable and robust gapped ($\pi,0$) SDW state at exactly half-filling. Furthermore, the magnetic, electronic, and orbital properties of this model showed that the three key experimental properties $-$ electronic structure, magnetic excitations, and ferro orbital ordering $-$ can be well understood within a single theoretical framework, as summarized below.

The electronic FS structure in our three-orbital model at half-filling reproduces two key features obtained in LDA calculations and observed in ARPES experiments: two hole pockets around the center point and elliptical electron pockets around the $M$ point of the unfolded Brillouin zone. The elliptical electron pockets, corresponding to quasi-one dimensional motion, lead to a Van Hove peak in the DOS near the Fermi energy, as is indeed obtained in LDA calculations. These electron pockets are therefore also responsible for the strong F spin coupling in our model, which results in maximum spin-wave energy at the FZB wave vector ${\bf q}=(\pi,\pi)$ as obtained in INS experiments. The strong F spin coupling in our model also accounts for the large planar anisotropy between $ab$ plane spin couplings as considered in phenomenological spin models.

Overall, the nature of magnetic excitations, spin-wave energy scale,  and absence of particle-hole continuum as calculated in our model are in agreement with experimental results. We also
explored the possibility of orbital ordering in our model due to lifting of degeneracy between the $d_{xz}$ and $d_{yz}$ bands and found the existence of a ferro orbital order between the $d_{xz}$ and $d_{yz}$ orbitals in the $(\pi,0)$ SDW state, in agreement with experiments.

\section*{Acknowledgements}
Sayandip Ghosh acknowledges financial support from Council of Scientific and Industrial Research, India.

\appendix
\section{Spin excitations in multi-orbital model}

We consider a general multi-orbital ($\pi,0$) ordered magnetic state involving ${\cal N}$ orbitals per site, and briefly discuss spin wave excitations in this state. The transverse spin susceptibility in this spontaneously broken-symmetry state is obtained from the time-ordered propagator of the transverse spin operators $S_{i \mu} ^-$ and $S_{j \nu} ^+$ for sites $i$, $j$ and orbitals $\mu$, $\nu$:
\begin{equation}
[\chi^{-+}(\omega)]_{i\mu;j\nu} = \int
dt e^{i\omega(t-t')}  \langle  T_t  S_{i\mu} ^- (t)
S_{j\nu} ^+ (t')\rangle, 
\end{equation}
where $T_t$ is the time ordering operator. As translational symmetry is preserved within the two-sublattice basis, Fourier transformation yields
\begin{eqnarray}
[\chi^{-+}({\bf q},\omega)]_{\alpha \beta} = \sum_{i} e^{-i{\bf q}\cdot ({\bf r}_i - {\bf r}_j)} [\chi^{-+}(\omega)]_{i\mu;j\nu},
\end{eqnarray}
where $\alpha$, $\beta$ refer to indices in the composite sublattice-orbital basis, and run through $1-2{\cal N}$.  

Retaining only ladder diagrams yields the spin wave propagator in the random phase approximation (RPA):
\begin{equation}
[\chi^{-+} _{\rm RPA} ({\bf q},\omega)] = \frac{[\chi^0 ({\bf q},\omega)]}{{\bf
1} - [U][\chi^0 ({\bf q},\omega)]}
\label{chi}
\end{equation}
expressed in a matrix form in the composite sublattice-orbital basis. Here,
the interaction matrix [U] includes $U$ (intra-orbital interaction) as diagonal elements and $J$ (Hund's coupling) as off-diagonal elements. The inter-orbital density interaction and the pair hopping terms do not contribute to magnetism up to RPA level.

The bare susceptibility is calculated from the Hartree-Fock (HF) level Green's function as: 
\begin{eqnarray}
[\chi^0 ({\bf q},\omega)]_{\alpha \beta} 
&=& i \int \frac{d\omega'}{2\pi} \sum_{\bf k} [G^{0}_{\uparrow}({\bf k},\omega')]_{\alpha \beta} \times \nonumber \\ 
&& [G^{0}_{\downarrow}({\bf k-q},\omega'-\omega)]_{\beta \alpha} \nonumber \\
&=& \sum_{{\bf k},l,m} \bigg[
{\frac{{\phi^{\alpha}_{{\bf k} \uparrow l}}{\phi^{\beta}_{{\bf k} \uparrow
l}}{\phi^{\alpha}_{{\bf {k-q}} \downarrow m}}{\phi^{\beta}_{{\bf {k-q}}
\downarrow m}}}{E^+_{{\bf {k-q}} \downarrow m} - E^-_{{\bf k} \uparrow l} +
\omega -i\eta}}  \nonumber \\
&+& {\frac{{\phi^{\alpha}_{{\bf k} \uparrow
l}}{\phi^{\beta}_{{\bf k} \uparrow l}}{\phi^{\alpha}_{{\bf {k-q}} \downarrow
m}}{\phi^{\beta}_{{\bf {k-q}} \downarrow m}}}{E^+_{{\bf k} \uparrow l} -
E^-_{{\bf {k-q}} \downarrow m} - \omega -i\eta}} \bigg] \\
\nonumber
\label{propagator}
\end{eqnarray}
in the sublattice-orbital basis, and involves integrating out the fermions in the ($\pi,0$) ordered spontaneously-broken-symmetry state. Here $E_{{\bf k}\sigma}$ and $\phi_{{\bf k}\sigma}$ are the eigenvalues and eigenvectors of the HF level Hamiltonian matrix and $l,m$ indicate the eigenvalue branches. The superscripts $+ (-)$ refer to particle (hole) energies above (below) the Fermi energy, and both inter-band and intra-band particle-hole terms are included. The spin wave energies are obtained from the poles of Eqn. \ref{chi}. The calculations  are performed over a 100 $\times$ 100  ${\bf k}$-point mesh and finite damping $\eta$ of 5 meV. Smaller values of $\eta$ and finer ${\bf k}$-point meshes do not produce qualitative or significant quantitative changes in our results.

\section{Hartree-Fock approximation}
Here, we discuss the Hartree-Fock (mean field) treatment of the onsite interaction terms given in Eq. \ref{interaction}. The mean-field parameters $n_{\mu}$ and $m_{\mu}$ describe the charge density and sublattice magnetization of the orbital $\mu$, and the rest of the notation is standard. The intra-orbital Coulomb interaction term in this approximation is given by:
\begin{eqnarray}
H_{\rm HF}^{\rm intra} &=& \frac{U}{2} \sum_{{\bf k}, \mu, \sigma} \bigg(-\sigma \tau m_{\mu}  + n_{\mu}\bigg) a_{{\bf k},\mu,\sigma}^\dagger a_{{\bf k},\mu,\sigma} \nonumber \\
&+& \text{const},
\end{eqnarray}
where $\tau$ = +(-) for sublattice A(B) in which $\uparrow$ ($\downarrow$) spins are majority. The inter-orbital Coulomb interaction and and the Hund's coupling terms take the forms:
\begin{eqnarray}
H_{\rm HF}^{\rm inter} = (U' - \frac{J}{2}) \sum_{{\bf k}, \mu \neq \nu, \sigma}  n_{\nu}  a_{{\bf k},\mu,\sigma}^\dagger a_{{\bf k},\mu,\sigma} + \text{const},
\end{eqnarray}
and 
\begin{eqnarray}
H_{\rm HF}^{\rm Hunds} = - \frac{J}{2} \sum_{{\bf k}, \mu \neq \nu, \sigma} \sigma \tau m_{\nu}  a_{{\bf k},\mu,\sigma}^\dagger a_{{\bf k},\mu,\sigma} + \text{const}
\end{eqnarray}
respectively. The pair-hopping term does not contribute to the mean-field level Hamiltonian.

The total interaction Hamiltonian decouples into magnetic and density terms:
\begin{eqnarray}
H_{\rm HF}^{I} &=&  - \sigma \tau \sum_{{\bf k}, \mu, \sigma}  \Delta_{\mu}  a_{{\bf k},\mu,\sigma}^\dagger a_{{\bf k},\mu,\sigma} \nonumber \\
&+& \frac{5J-U}{2} \sum_{{\bf k}, \mu, \sigma} n_{\mu}a_{{\bf k},\mu,\sigma}^\dagger a_{{\bf k},\mu,\sigma} + \text{const},
\end{eqnarray}
with the self-consistent exchange fields defined as $2\Delta_\mu = U m_{\mu} + J\sum_{\nu \neq \mu}m_{\nu}$. Here, we have taken a constant total electron filling ($\sum_\mu n_\mu$) and $U' = U - 2J$.

\end{document}